\documentclass{article}[14pt]
\usepackage{blindtext}
\usepackage{geometry}
\usepackage[utf8]{inputenc}
\usepackage[T1]{fontenc}
\usepackage[english]{babel}
\usepackage{authblk}
\usepackage{cancel}
\usepackage{amsmath}
\usepackage{amssymb}
\usepackage{amsfonts}
\usepackage{graphicx}
\usepackage{comment}
\usepackage{float}
\usepackage{xcolor}
\usepackage[normalem]{ulem}
\usepackage{caption}
\usepackage{array}
\usepackage{setspace}
\usepackage{subcaption}
\usepackage{booktabs}
\usepackage{xcolor}
\usepackage[tableposition=above]{caption}
\usepackage{soul}
\usepackage{pifont}
\usepackage{adjustbox}
\usepackage{multirow}
\usepackage{abstract}
\usepackage{bm}

\usepackage{hyperref}
\usepackage{cleveref}

\usepackage{csquotes}
\usepackage[style=authoryear,natbib=true,maxbibnames=99,doi=false]{biblatex}
\renewbibmacro{in:}{}
\DeclareNameAlias{sortname}{family-given}

\begin{document}

\singlespacing
\font\myfontA=cmr12 at 16pt
\font\myfontB=cmr12 at 12pt
\title{\myfontA Heterogeneous Image-based Classification Using Distributional Data Analysis  \\ \vspace{.25cm}}

\author{\myfontB Alec Reinhardt$^1$\footnote{{\it Corresponding author}: Alec Reinhardt, Department of Biostatistics, UT MD Anderson Cancer Center, 1400 Pressler St., Floor 4, FCT4.6000, Houston, TX 77030. {\it E-mail}: aereinhardt@mdanderson.org}, Newsha Nikzad$^2$, Raven J. Hollis$^2$, Galia Jacobson$^2$, Millicent A Roach$^2$, Mohamed Badawy$^3$, Peter Chul Park$^4$, Laura Beretta$^5$,  Prasun K Jalal$^6$, David T. Fuentes$^7$, Eugene J. Koay$^{2,8}$, and Suprateek Kundu$^1$ \\
{$^1$Department of Biostatistics, MD Anderson Cancer Center} \\
{$^2$ Cancer Physics and Engineering Laboratory, MD Anderson Cancer Center} \\
{$^3$Department of Pediatrics, UT Southwestern Medical Center} \\
{$^4$Department of Radiation Oncology, UC Davis Comprehensive Cancer Center} \\
{$^5$Department of Molecular and Cellular Oncology, MD Anderson Cancer Center}\\
{$^6$Department of Hepatology, Baylor College of Medicine}\\
{$^7$Department of Imaging Physics, MD Anderson Cancer Center}\\
{$^8$Department of Radiation Oncology, MD Anderson Cancer Center}
}

\date{\today}
\maketitle

{\centerline{\bf{Abstract}}}

Diagnostic imaging has gained prominence as potential biomarkers for early detection and diagnosis in a diverse array of disorders including cancer. However, existing methods routinely face challenges arising from various factors such as image heterogeneity. We develop a novel imaging-based distributional data analysis (DDA) approach that incorporates the probability (quantile) distribution of the pixel-level features as covariates. The proposed approach uses a smoothed quantile distribution (via a suitable basis representation) as functional predictors in a scalar-on-functional quantile regression model. Some distinctive features of the proposed approach include the ability to: (i) account for heterogeneity within the image; (ii) incorporate granular information spanning the entire distribution; and (iii) tackle variability in image sizes for unregistered images in cancer applications. Our primary goal is risk prediction in Hepatocellular carcinoma that is achieved via predicting the change in tumor grades at post-diagnostic visits using pre-diagnostic enhancement pattern mapping (EPM) images of the liver. Along the way, the proposed DDA approach is also used for case versus control diagnosis and risk stratification objectives. Our analysis reveals that when coupled with global structural radiomics features derived from the corresponding T1-MRI scans, the proposed smoothed quantile distributions derived from EPM images showed considerable improvements in sensitivity and comparable specificity in contrast to classification based on routinely used  summary measures that do not account for image heterogeneity. Given that there are limited predictive modeling approaches based on heterogeneous images in cancer, the proposed method is expected to provide considerable advantages in image-based early detection and risk prediction.

\vskip 5pt


\section{Introduction}
\label{sec:introduction}
Hepatocellular carcinoma (HCC) is the most common type of primary liver cancer and a significant global health concern. It is the sixth most common cancer and the fourth leading cause of cancer-related deaths worldwide. One of the key factors contributing to its high mortality rate is the lack of early detection. Early detection plays a pivotal role in improving liver cancer prognosis and increasing treatment success rates. Patients diagnosed at an early stage have the opportunity to undergo curative treatments such as resection, ablative therapies, and liver transplantation, while those diagnosed at a later stage typically only qualify for palliative systemic treatments with limited effectiveness. Consequently, the 5-year survival rate surpasses 70\% for individuals with early-stage HCC, whereas it falls below 5\% for those diagnosed at advanced stages. Surveillance for HCC is recommended in at-risk patients, including those with cirrhosis of all etiologies, and certain populations with chronic HBV infection. Guidelines recommend HCC surveillance with abdominal ultrasound (US) with or without serum alpha-fetoprotein (AFP) measurement \citep{yildirim2023advances} every 6 months. Other diagnostic imaging modalities for surveillance include computed tomography (CT) and magnetic resonance imaging (MRI), and ultrasound (US) imaging. However, the sensitivity of the surveillance may potentially vary with the imaging modality \citep{tzartzeva2018surveillance}.

Although medical imaging has made rapid advances and has emerged as one of the most promising tools for early diagnosis and risk detection, there are important limitations. These include both MRI-based over surveillance due to the frequent detection of benign vascular lesions, and delayed HCC diagnosis due to the under-staging of small or early malignant lesions in approximately 25\% of cases \citep{mahmud2020risk, park2020enhancement}. A potential reason for these inadequacies is the inherent noise in liver MRI scans, which is often unaccounted for. Another major diagnostic and prognostic challenge of rising concern is the heterogeneity of pre-cancerous and cancerous lesions. Despite technological advances, lesion detection, characterization of patterns of enhancement, and monitoring remain sub-optimal. While the Liver Imaging Reporting and Data System (LI-RADS) has been developed as a standardized classification system for liver nodules (depending on lesion size, arterial phase enhancement, capsular enhancement, washout, and threshold growth to deduce a nodule’s probability of malignancy, and so on), risk stratification is hindered by significant heterogeneity within LI-RADS categories, especially for LR-3 and LR-4 lesions of which 38\% and 74\% are HCC, respectively \citep{chernyak2018liver, nikzad2022characterization}. A recent meta-analysis showed that there was significant heterogeneity in sensitivity of US-based surveillance (21\%-89\%) across studies included in the meta-analysis \citep{tzartzeva2018surveillance}, which can be potentially attributed to the above factors and highlights a limitation of imaging-based surveillance. Fuethermore, in other domains such as pancreatic cancer, standardized scoring system such as LI-RADS may not always be available. This exacerbates difficulties for imaging-based risk stratification, and the standard of care may resort to biopsy or other invasive procedures for a definitive diagnostic assessment. Therefore, there is critical need for new minimally invasive imaging platforms and accompanying statistical approaches that can render risk stratification with high sensitivity and specificity based on imaging data collected at earlier disease stages.


In this article, we propose novel statistical and machine learning models for early detection and risk prediction in cancer imaging studies, motivated by HCC application. In particular, our goal is to develop a novel distributional data analysis (DDA) based approach leveraging quantile distributions \citep{matabuena2021distributional} across imaging pixels to be used for  classification or prediction. This article makes several important contributions. 

First, we use a novel type of liver image known as enhancement pattern mapping (EPM) that are derived from T1-MRI scans and have higher contrast-to-noise ratio. EPM is a novel pixel-based signal analysis technique that quantifies the difference in enhancement over time of a given pixel in the liver compared to either a patient-specific or population-based normal liver model, providing a measurement and visualization of how different the signal is over an entire volume of interest. Previous literature \citep{park2020enhancement} has shown that EPM derived from CT images improves the contrast-to-noise (CNR) ratio enhancement for lesion detection in hepatobiliary malignancy. In contrast to this previous research, we use EPM features derived from multi-phase MRI scans in order to evaluate the predictive potential of a novel image-based biomarker. Secondly from a methodological perspective, we propose a novel distributional data analysis (DDA) approach that uses the entire quantile distribution of pixel-level imaging measurements for classification and risk prediction. Such an approach extends the DDA literature proposed for regression \citep{yang2020quantile} to cancer imaging  involving high-dimensional heterogeneous images. By utilizing full information embedded in the probability distribution of granular pixel-level measurements, the proposed approach is able to account for the heterogeneity within the image that may arise from the differences of histopathological properties of lesional, peri-lesional and healthy tissues. Therefore, this approach provides considerable advantages over routinely used approaches based on summary measures. Another important advantage of the DDA based method is that one does not require the images to be registered across samples and the methodology can handle lesions of varying sizes that is often encountered in cancer imaging applications. Third, we develop a novel scalar-on-functional quantile classification methodology that leverages penalization techniques to achieve sparsity and model parsimony in the selected features. Fourth, we focus on several important analysis goals including case vs control diagnosis and risk stratification, along with the critical goal of {\it early detection} that involves identifying lesions presenting the highest risk of longitudinal progression based on data from the initial disease stages. These goals are described in more detail below. 

 Our initial objective involves disease classification, aiming to distinguish between cases and controls using whole liver images with and without lesion mask information. The subsequent task is risk stratification, focusing on classifying the grade of HCC liver lesions (mild versus aggressive). Following this, we shift our attention to the primary objective of early detection, which holds significant clinical and translational importance by identifying lesions with high potential for longitudinal progression early on, thus improving survival outcomes. Specifically, in aim 3, we employ the proposed scalar-on-functional quantile classification approach to identify lesions likely to exhibit aggressive longitudinal progression based on imaging data from the pre-diagnostic visit. We predict the tumor grade at the post-diagnostic visit using training images from the pre-diagnostic visit and evaluate performance accordingly. In the above classification analyses, we compare the performance of different combinations of functional quantile distributional features and structural radiomics features as well as other image summaries. Collectively, these combined analyses establish a comprehensive prediction framework based on pixel-level heterogeneous imaging data, capable of addressing gaps in literature. In particular, existing approaches primarily rely on summary imaging features without accounting for heterogeneity with the image, resulting in pitfalls.

\begin{figure}[h]
    \centering
    \includegraphics[width=\textwidth]{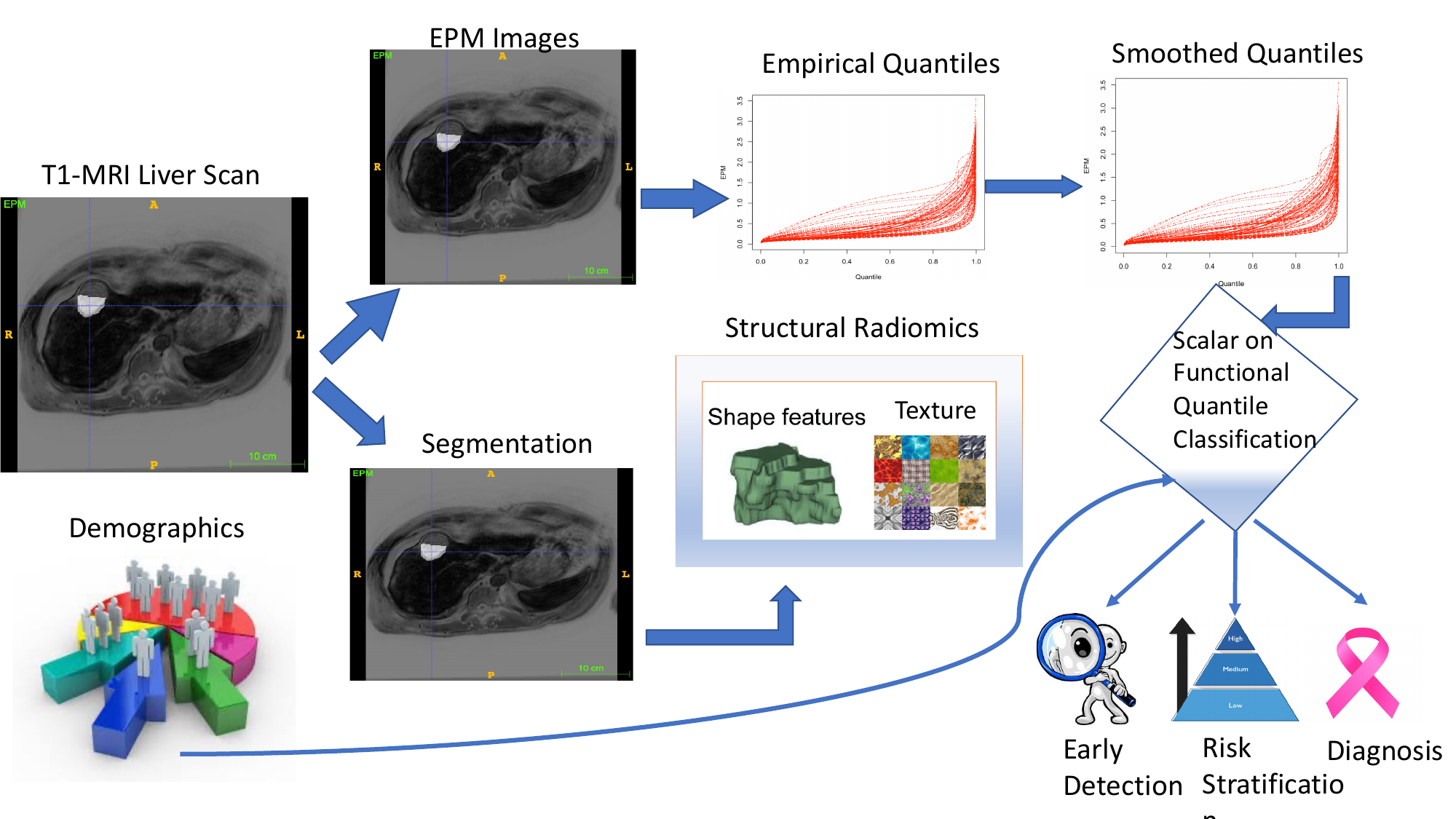}
    \caption{A schematic of the proposed analytical pipeline. The leftmost panel captures the 5 different phases of the T1-MRI liver scan that are all registered to a common template for a given subject. Subsequently the EPM images are computed from these T1-MRI scans, and then empirical quantile distributions followed by smoothed quantile distributions (based on quantlet basis expansions) were computed. The arterial phase of T1-MRI is used for segmentation of tumor and also used to compute the structural radiomics features. These two types of features, along with demographics are used for three classification tasks that include early detection, risk stratification, and diagnosis (case vs controls).}
    \label{fig:schema}
\end{figure}

Furthermore, we conduct thorough numerical comparisons demonstrating the notable benefits of integrating functional quantile distribution features with structural radiomics features for classification. This surpasses the effectiveness of using functional quantile feature or structural radiomics features by themselves or even routinely used summary measures derived from the image. Our meticulous numerical experiments reveal that even with a limited to moderate number of functional quantlet features in the classification model, the proposed method achieves sensitivity levels exceeding 80\% and specificity nearing 95\% for early detection. Additionally, we observe consistent performance across varying lesion sizes and despite small to moderate training sample sizes. These findings clearly underscore the potential of the proposed functional quantile distributions extracted from EPM images in cancer diagnosis and risk prediction, especially when combined with structural radiomics features. A schematic diagram of the proposed analytical pipeline is presented in Figure \ref{fig:schema}.


\section{Related Literature}
There is a growing statistical and machine learning literature for image-based modeling as summarized below. 

\vskip 10pt

{\noindent \underline{Analysis using vectorized pixels:}} The earliest approaches for image-based analysis relied on mass univariate regression analysis \citep{friston1995characterizing} that may not account for dependence across imaging pixels. Later, more sophisticated Markov Random Field (MRF) pixel-wise regression approaches have been proposed that are able to account for the spatial dependence \citep{smith2007spatial, lee2014spatial}. While useful, pixel-level approaches described above often encounter the curse of dimensionality due to tens of thousands of imaging pixels in a single image. Recently, tensor-based approaches have been proposed \citep{kundu2023bayesian, guhaniyogi2017bayesian} that simultaneously  alleviate the curse of dimensionality and  account for the spatial information in the images. However, the above approaches were often designed for neuroimaging data where the images are required to be registered across subjects. While registration of brain images is standard in neuroimaging literature, established pipelines for image registration across samples are typically not readily available for cancer imaging studies due to various factors such as varying size, composition, and other characteristics of the organs of interest. 
Therefore, the above sophisticated methods are not directly generalizable to cancer imaging applications that is our focus.

\vskip 10pt

{\noindent \underline{Analysis based on summary radiomics features:}} Summary features such as mean pixel intensity \citep{Avulamean} as well as more informative radiomic features are extracted from high-dimensional radiological images \citep{ lambin2012radiomics} via established algorithms. These features capture different global features of the image such as first order and second order statistics, as well as image texture. Radiomic analyses have been applied within oncology to improve diagnosis and prognosis  \citep{lambin2017radiomics,aerts2014decoding} 
with the aim of delivering precision medicine. The radiomic workflow involves curation of clinical and imaging data and is a step-wise process involving image pre-processing, tumor segmentation, feature extraction, model development, and validation \citep{larue2017quantitative}. Such a pipeline requires input from individuals in many disciplines and at multiple levels, resulting in 
an overly supervised approach that is potentially susceptible to subject-specific bias. Moreover, radiomics features collapse information across the whole image or regions of interest (ROI) that leads to information loss, and they may fail to capture the heterogeneity within the image, which is a key (but often overlooked) concern  \citep{dagogo2018tumour}.

\vskip 10pt

{\noindent \underline{Scalar-on-function models:}} Methods for functional data analysis \citep{ramsay1991some} have become ubiquitous with the growth of recent technologies that are able to generate high-dimensional functional data. Although the vast majority of literature has focused on one-dimensional functional curves \citep{morris2015functional}, recent literature has started investigating models involving more complex types of functional data such as images. For example, neuroimaging analysis using entire brain images as covariates (scalar-on-image regression) for prediction of a continuous outcome have gained in popularity \citep{feng2020bayesian, ma2022multi}. In these types of models, the image is treated as a functional object.  Typical scalar-on-image regression approaches account for hundreds of thousands of pixels via low-dimensional representation of images, for instance principal components or tensor methods.  
For example, approaches involving functional principal component analysis (FPCA) \citep{zipunnikov2011functional} often assume that the components driving variability in the images are related to the outcome that may not always be practical. 
An advantage of these dimension reduction methods is that they do not require the image to be registered across samples. However, FPCA approaches only use a subset of principal components resulting in information loss and potentially compromising predictive accuracy. Alternative approaches involving basis expansion representations have been proposed for scalar-on-image regression \citep{wang2017generalized, reiss2015wavelet} and provide a desirable avenue to preserve the spatial properties of the image. Unfortunately, such basis expansion methods typically require images to be registered across samples that may not feasible in cancer imaging studies. Moreover, basis expansion approaches typically do not have in-built mechanisms for low dimensional feature extraction and therefore may suffer from the curse of dimensionality.

\vskip 10pt

{\noindent \underline{Statistical literature on distributional data analysis:}} Distributional data analysis (DDA) can be considered a specific form of functional data analysis, where the covariate probability (or quantile) distribution is used as the functional predictor. This type of approach requires repeated measures from a given observed covariate, which is increasingly common in recent studies with growing volumes of data. In such studies, each covariate in a data set is associated with its own probability distribution that represents the heterogeneity in the features of interest, which is subsequently used in modeling approaches. Such studies may involve user-defined histograms to be used as predictors in scalar-on-function regression \citep{augustin2017modelling}, or compositional data analysis (CoDA) \citep{dumuid2020compositional} that refers to another group of methods to model continuously measured wearable data to analyze samples of densities \citep{talska2021compositional}. Distribution-on-distribution regression models were suggested by \citep{ghodrati2022distribution}. 
Additional work on DDA involve research by  \citep{ghosal2023distributional} who used subject-specific quantile functions to capture the distributional nature of wearable data, and  \citep{matabuena2021distributional} who used a quantile-function representation of accelerometry data to predict health outcomes. While the use of DDA approaches is growing rapidly, these techniques have not yet been adopted for mainstream cancer imaging applications accounting for heterogeneity, to our knowledge. 

\section{Methods and Materials}

\subsection{Description of Liver Imaging Dataset}

The study population included patients with cirrhosis presenting to the Hepatology Clinic between 2012 and 2020 at a single tertiary care hospital who were prospectively followed. Patients underwent HCC surveillance with the instructional practice of contrast-enhanced MRI every six months. Patients with cirrhosis were included in the study if they had at least two consecutive contrast-enhanced MRIs, utilizing liver protocol for HCC surveillance. Patients with cirrhosis and HCC diagnosed from baseline imaging were excluded. The dataset subsequently comprised of {\color{black} 48} individuals with LR-3 and LR-4 liver lesions (cases), and 97 healthy subjects without any lesions (controls). {\color{black} Among the case subjects, we excluded certain subjects and/or visits based on quality control for the EPM images and lesion segmentations. After exclusion, imaging and clinical data on 38 case subjects were available for analysis, out of which 35 subjects had imaging data for both timepoints.} Among the {\color{black} 73} case images (35 subject $\times$ 2 visits + 3 subjects $\times$ 1 visit), {\color{black}91} distinct lesion areas were identified including multiple lesions for a {\color{black} subset} of subjects. These 91 lesions include all pre-diagnostic lesions and post-diagnostic lesions but exclude minuscule-sized lesions with about 10 or less pixels. Longitudinal differences in tumor characteristics and grades were observed between the pre-diagnostic and post-diagnostic images for 35 subjects with imaging data for both timepoints.
In addition to the MRI scans and corresponding lesion and liver masks, the LIRADS score for each lesion were also available that ranged from 2-5. Arterial phase hyperenhancement of the lesion, venous and delayed phases washout, in addition to the lesion’s size and growth pattern were the criteria used to assign category codes based on LI-RADS version 2018 guidelines.  Spatially distinct lesioned areas are identified using a clustering algorithm. 
When multiple distinct lesioned areas are identified, we separate them for processing when the nearest points lie at least 5 mm apart. Figure \ref{fig:Fig2} provides visual illustrations. 

\subsubsection{Enhancement Pattern Mapping (EPM) Images}
The original T1w-MRI scans were processed to obtain voxel-wise EPM data that corresponds to the blood perfusion in the liver tissue.  EPM is a three-dimensional, voxel-based algorithm that was earlier used for quantitative image analysis of liver lesions based on CT scans \citep{park2020enhancement}, which we now adapt for MRI scans. The generalized enhancement pattern, such as quantitative intensity changes throughout multi-phase MRI due to uptake and washout of contrast materials was acquired from the registered multi-phase MRI scans. A normal liver enhancement curve was developed through fitting the MRI intensity values over time within user-selected ROIs, sampled uniformly across cirrhotic liver parenchyma from the same patient. The root-mean-square deviation (RMSD) for each voxel was computed by averaging the squares of differences between the generalized normal liver intensity and the voxel intensity across all time points, and subsequently taking the square root of the average. The RMSD values of each voxel was mapped to their original MRI coordinates, and these values comprised the EPM image maps used for our analysis. The normal liver enhancement curve was obtained by fitting MRI intensity values sampled from the normal liver ROIs over the period of contrast phases by a piece-wise smooth function, where each piece was a second-order polynomial. The EPM algorithm was implemented numerically using MATLAB \citep{MATLAB}.


\subsection{Extraction of radiomics features} 

We engineer and extract novel functional radiomics features from the EPM images, and combine them with standard structural radiomics features from the MRI scans in our analysis. 

\vskip 10pt

{\noindent \underline{Functional Quantlet Features based on EPM}:}  We computed functional quantlet features that capture the information embedded in the distribution of EPM values across all pixels in the lesion and/or the liver (depending on the analysis goals). These functional features encompass a much richer set of information spanning the entire distribution compared to summary statistics typically reported in existing pipelines such as \citep{van2017computational}. The functional quantlet features rely on a basis expansion that enables smoothing and potential interpolation, thus reducing the impact of unequal-sized lesions. Further, they enable a parsimonious representation via a near-lossless property that ensures that only a moderate number of these quantlets are able to capture almost all of the information in the quantile distribution. 
The full mathematical details for deriving the quantlet features are described in the next section.

\vskip 10pt

{\noindent \underline{Structural radiomics features based on T1-MRI:}}   In addition to EPM images, structural radiomics features were also extracted based on the arterial phase of MRI scans. For controls, each ROI (6mm in diameter) was manually defined over background liver parenchyma, and subsequently the structural radiomics features were extracted from these ROIs using the {\it Pyradiomics} package \citep{van2017computational}. For cases, the structural radiomics features were computed separately for each lesion as well as for an independent non-lesional (pre-defined) ROI. Both liver masks and lesion masks were used for defining ROIs, where the lesion segmentation was conducted based on the arterial phase of the T1w-MRI scan. When computing the structural radiomics features for cases, we did not differentiate between multiple lesions with the same LIRADS score.

From a total of 108 radiomics features extracted from suitable ROIs, we included four categories of structural radiomics information: 24 Gray Level Co-occurrence Matrix (GLCM) features, 16 Gray Level Run Length Matrix
(GLRLM) features, 16  Gray Level Size Zone Matrix (GLSZM) features, and 14 shape-based features. We excluded variables representing first-order structural radiomic information (e.g. energy, entropy, interquartile range) due to potential overlap with functional radiomic information. 


\subsection{Methodology for extracting functional radiomics features via quantlet basis approach}
\begin{figure}
    \centering
    \includegraphics[width=.65\textwidth]{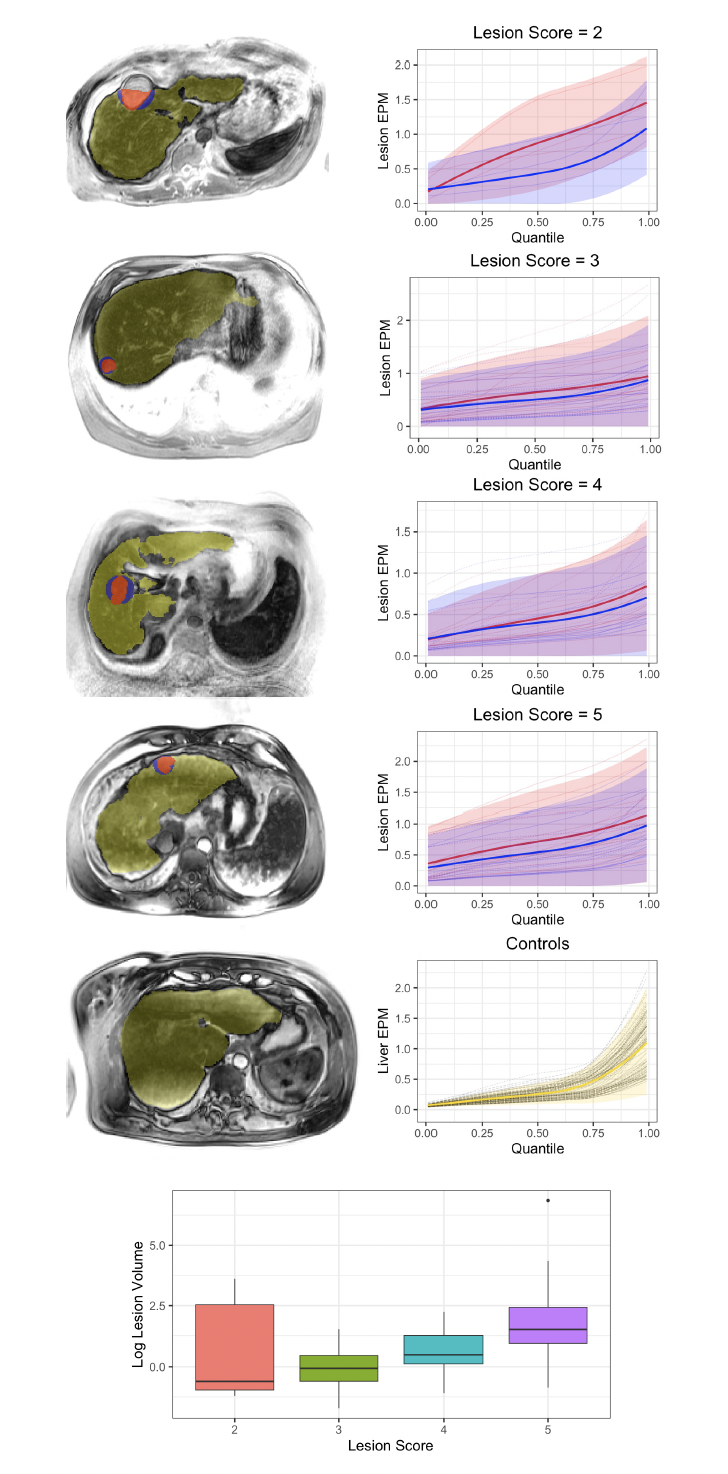}
    \caption{\label{fig:Fig2} Visualizations of lesion EPM images with segmentations for lesion grades 2-5 and healthy control. Liver masks are shown in yellow; lesion masks in red and surrounding peri-lesional areas are shaded in blue. Mean empirical quantiles are shown along with 95\% confidence bands for each lesion grade and controls. For cases, quantiles are shown across all lesional (red) and peri-lesional (blue) areas, and for controls, quantiles are shown across all liver masks (yellow). Distributions of log-lesion volumes are also shown for each lesion grade. Observed lesion volumes ranged from 0.17-0.950 cm$^3$, with approximate diameters of 0.70-12.2 cm.}
\end{figure}

The starting point for the proposed scalar-on-functional quantile classification approach is to compute the empirical quantiles. The quantile distributions naturally account for heterogeneity in the image (arising from lesional, peri-lesional, and healthy regions in the organ) and is able to tackle different image sizes across samples. Figure \ref{fig:Fig2} visualizes differences in the quantile distributions between different lesion grades as well as between cases vs controls. It is clear that the quantile distributions for the lesion pixels are often flatter compared to the distributions for non-lesional pixels. In particular, the non-lesional distribution has near-zero values and higher values corresponding to left and right tails respectively, which represents greater variability in the pixel distribution. 


First, we demonstrate how to derive the empirical quantiles from the EPM images. 
Denote the $i$th image as $X_i$ comprising $m_i$ pixels ($i=1,\ldots,N$). We note that $X_i$ may include the full liver image for subject level classification and a smaller image involving lesional and per-lesional areas for lesion level classification. Further, denote the $m_i$ empirical quantiles for the $i$th sample as $\{Q_i(p_{1}),\ldots,Q_i(p_{m_i})\}$, which are derived from $X_i$, $i=1,\ldots,N,$ for $0<p_j<1$. Mathematically, the quantile function is defined as $Q_X(p) = F^{-1}_X(p) = inf\{x: F_X(x)\ge p \}, \mbox{ } 0\le p \le 1, $
where $inf(\cdot)$ denotes the infimum, and $F_X$ denotes the cumulative distribution function, i.e. $p=F_X(x) = P(X\le x)\in [0,1]$. Further, denote the order statistics from the $i$th image as $\{X^{(1)}_i, \ldots, X^{(m_i)}_i \}$ that are arranged in an increasing order. We note that for $p\in [1/(m_i+1), m_i/(m_i+1) ]$, the empirical quantile function for the $i$th sample is computed as 
\begin{eqnarray}
\hat{Q}_i(p) = (1-w_i)X_i^{((m_i+1)p)} + w_i X_i^{((m_i+1)p+1)}, \label{eq:emp_quantile}
\end{eqnarray}
where $w_i = X_i^{((m_i+1)p+1)}- X_i^{(m_i+1)p}$. The above expression is obtained via interpolation, where $[x]$ is an integer $\le x$. In practice, we will focus our modeling on quantiles lying within $[\delta,1-\delta]$ where $\delta = \max_{i\le n} \{ 1/(m_i+1) \}$.

We note that the empirical quantile function can be computed on a grid $\{p_1,\ldots,p_K\}$ lying within $[\delta,1-\delta]$ using (\ref{eq:emp_quantile}), where the grid is standardized across images. Subsequently, the corresponding empirical quantiles $\{Q_X(p_1),\ldots,Q_X(p_K) \}$ could be potentially used as covariate features for prediction. However, the quality of the empirical quantiles computed over the standarized grid will depend on the size of the images. While it may be possible to estimate these quantiles with reasonable accuracy corresponding to larger images with several thousands of pixels, the same may not be true for smaller images with fewer pixels or small lesions. This is due to the fact that the order statistics $\{X^{(1)}_i, \ldots, X^{(m_i)}_i \}$ are potentially spaced farther apart for images with fewer pixels, that results in an inferior interpolation in (\ref{eq:emp_quantile}). Therefore, directly using the empirical quantile distribution as functional predictors may not be desirable, particularly for lesion-level analysis. 


While useful, it may not be appealing to directly use the empirical quantiles in the modeling approach due to the fact that their accuracy depends on the image size and other factors such as lack of smoothness across neighboring quantiles. In particular, the empirical quantile estimates may be sensitive to the number of pixels in the image due to inadequate interpolation. A more appealing approach is to propose a statistical smoothing method over the empirical quantiles with the goal being to provide more accurate and robust quantile-based features. As shown in \cite{yang2020quantile}, it is possible to find an empirical set of basis functions $\{ \psi_1, \ldots, \psi_K\}$ on the quantile space to approximate the empirical quantiles with extremely high accuracy (a property known as near lossless). That is,  
\begin{eqnarray}
Q_{X_i}(p_j) \approx \sum_{k=1}^K Q^*_{ik} \psi_k(p_j), \label{eq:basis_expansion}
\end{eqnarray}
where $Q^*_{ik}$ represent the basis coefficients for the $k$th basis function for the $i$th sample, and $\{ \psi_1(p),\ldots, \psi_K(p)\}$ represent a set of orthogonal quantlet basis functions. Under suitable choice of basis functions, even a small to moderate number of basis functions in (\ref{eq:basis_expansion}) is sufficient to guarantee an accurate approximation, as shown in \cite{yang2020quantile}. 
In contrast to the empirical quantiles that is derived directly from the imaging data, the functional quantlet approach uses a wavelet-based smoothing approach that leads to information sharing across neighboring quantile levels. Therefore, empirical quantiles are more susceptible to noise in the image and the imaging size compared to the functional quantlet-based approach.

The basis coefficients in (\ref{eq:basis_expansion}) are obtained in a straightforward manner as follows. Let us denote $\boldsymbol {Q_i} = (Q_i(1),\ldots,Q_i(m_i))$, and $\boldsymbol\Psi $ be a $K\times m_i$ with the $\boldsymbol\Psi(k,j) = \psi_k(p_j)$. Then the basis coefficients are given as $\boldsymbol  Q^*_i = (Q^*_{i1}, \ldots, Q^*_{iK}) = \boldsymbol Q_i \boldsymbol\Psi^T_i (\boldsymbol\Psi_i \boldsymbol\Psi^T_i)^{-1}$, where $K<< \min\{m_1,\ldots, m_N\}$ in our settings involving a large number of pixels in the image. We note $\boldsymbol Q^*_i$ contains virtually all of the information in the raw data $\boldsymbol Q_i$ via the near lossless property, but it is simultaneously more robust to noise and overfitting in the empirical quantile distribution. Therefore, we use the smoothed quantile function $\hat{Q}_{X_i}(p_j) = \sum_{k=1}^K \hat{Q}^*_{ik} \psi_k(p_j)$ as the functional predictor in our model in lieu of the empirical quantile distribution, where $\hat{Q}^*$ denotes the estimated coefficients under (\ref{eq:basis_expansion}). More details on such coefficient estimation as well as the choice of $K$ are provided in the following subsection. 

\subsection{Choice of Quantlet basis}

We adapt the approach in \cite{yang2020quantile} for choosing the basis functions, and design quantlets as a suitable set of basis functions that were shown to work well for representing the quantile distribution. We create empirical quantlet bases from a given dataset, serving as a robust representation of each subject's quantile function. The basis function is able to reconstruct the quantile distribution with greater than $99.9\%$ accuracy using a moderate number coefficient terms. This type of low dimensional feature extraction results in massive dimension reduction for large images and yields a more faithful reconstruction of the quantile distribution compared to alternate representations such as Legendre-polynomials \citep{ghosal2023distributional} that can result in poor tail approximation for heavy tailed distributions \citep{hosking1990moments}. These bases are employed in our subsequent quantile functional regression model. 





Consider the set of quantile functions $\mathcal{Q}$ such that $\mathcal{Q}:= \{ Q:\int \big(Q^2(p) d\Pi(p) \big)^{1/2} < \infty \}$, where $\Pi(\cdot)$ is a uniform density with respect to the Lebesgue measure. Define the first two basis functions to be a constant basis $\zeta_1(p)=1$ for $p\in [0,1]$ and standard normal quantile function $\zeta_2(p)=\Phi^{-1}(p)$. These orthogonal bases cover the entirety of Gaussian quantile functions, where the initial coefficient signifies the mean and the second coefficient represents the distribution's variance. An overcomplete dictionary is created, comprising these bases as well as numerous elements generated from Beta cumulative density functions (CDFs). The additional basis functions are constructed as $\zeta_j(p) = P_{N^{\perp}}(\frac{F_{\theta_k}(p) - \mu_{\theta_k}}{\sigma_{\theta_k}}), j\ge 3,$
where $F_{\theta_k}$ is the cumulative distribution function (CDF) of a Beta$(\theta_k)$ distribution for some positive parameters $\theta_k = \{a_k, b_k \}$, $\mu_{\theta_k}=\int_{0}^1 F_{\theta_k}(u) du$, and $\sigma^2_{\theta_k} = \int_{0}^1 (F_{\theta_k}(u) - \mu_{\theta_k})^2 du$ are the centered and scaled values of these distributions for standardization, respectively, and $P_{N^{\perp}}$ indicates the projection operator onto the orthogonal complement to the Gaussian basis elements $\zeta_1$ and $\zeta_2$. In other words, $P_{N^{\perp}}(f(p)) = f(p) - \zeta_1(p) \int_{0}^1 f(u)\zeta_1(u) du - \zeta_2(p) \int_{0}^1 f(u)\zeta_2(u) du $.

Put together, the set $\mathcal{D}^O = \{ \zeta_1 \cup \zeta_2\} \cup \{ \zeta_k: \theta_k \in \mathcal{\theta} \}_{k=1}^{K_0}$ represents an overcomplete dictionary, where  $\mathcal{\theta} $ represents the parameter space of interest. In practice, to fix the number of dictionary elements, we choose a grid on the parameter space to obtain $\mathcal{\theta} = \{\theta_k = (a_k,b_k) \}_{k=3}^{K_0}$ by uniformly sampling on $\mathcal{\theta}\subset (0,J)^2$ for some sufficiently large $J$, and and choosing $K_0$ to be a large integer. More details on how to choose the parameter space is provided in \cite{yang2020quantile}, along with additional theoretical results that justify the choice of a large dictionary of Beta CDF to represent any quantile function whose first derivative is absolutely continuous. 

Once this overcomplete basis set has been constructed, a sparse subset of basis functions are chosen that can represent the quantile functions across all samples. This is made possible by first fitting a Lasso regression \citep{tibshirani1996regression} to select a sparse set of wavelets for each subject separately, and subsequently taking the union across these sparse sets of basis elements across all subjects, to obtain a unified set of dictionary elements $\mathcal{D}$. Subsequently, a leave-one-out concordance correlation coefficient (LOOCCC) is proposed that allows one to reduce the number of basis elements in $\mathcal{D}$ even further, with minimal loss of information. This property is known as near lossless property. This goodness of fit criteria for the $i$th subject is given as $\rho_{(i)} = \frac{Cov(Q_i(\cdot),\sum_{k\in \mathcal{D}_{(i)}^U} \zeta_k(\cdot) Q^U_{ik})}{Var(Q_i(\cdot)) + Var(\sum_{k\in \mathcal{D}_{(i)}^U} \zeta_k(\cdot) Q^U_{ik})} $, where $\mathcal{D}_{(i)}^U$ is the set of the sparse basis elements obtained under the Lasso regression, combined across all but the $i$-th subject. Subsequently one can use $\rho^0= \min \{ \rho_{(i)}\}_i$ as an overall goodness of fit measure that can be used to select a small to moderate number of basis elements for our downstream prediction model, with no more than minimal loss of information  For example, in the Liver EPM dataset, we find that $13$ basis elements are able to preserve a concordance of at least 0.999, {\color{black}when reconstructing the quantile distributions from the entire liver mask}. 
Once a reduced set of basis elements are obtained, subsequent Gram-Schimdt orthogonalization steps are used followed by a wavelet based denoising step \citep{donoho1995wavelet}.



\subsection{Classification based on extracted radiomics features}

\subsubsection{Proposed Model}
We generalize the classical logistic regression models \citep{pan2021safe} to include functional predictors in the form of smoothed quantile distributions. We use the derived functional quantlet features, along with the extracted structural radiomics features and demographic features for risk prediction and early diagnosis.  Let $y_i\sim Ber(p_i(Q_x))$ denote the binary outcome of interest, which follows a Bernoulli distribution with probability $p_i(Q_x)$ that depends on the quantile distribution of the image. Further, let ${\bf y} = (y_1,\ldots, y_N)$ denote the vector of binary outcomes, ${\bf w}$ denotes the demographic features, and ${\bf z}$ denotes the structural radiomics features derived from the arterial phase of the T1w-MRI scans. {\color{black} Let $\mathcal{X}
_i$ correspond to the full liver image for Aim 1A, and let it correspond to pre-defined ROIs encompassing the lesion for the cases in Aim 1B and for the lesion level analyses in Aims 2-3.}  When lesion segmentation is available, denote the lesion mask for the $i$th image as $\mathcal{L}_i$, so that $\mathcal{L}_i^c = X_i \setminus \mathcal{L}_i$ denotes the peri-lesional areas adjacent to the lesion. The likelihood is $L({\bf y}\mid \mathcal{X}) = \prod_{i=1}^N p^{y_i}_i(Q_x)(1-p_i(Q_x))^{1-y_i}$, where
\begin{align}
    & \frac{p_i(Q_x)}{1-p_i(Q_x)}  = \alpha + ({\bf w}_i^T, {\bf z}_i^T)\bm{\zeta} + \int_{0}^1 \hat{Q}_{\mathcal{X}_i}(p) \beta_{\mathcal{X}}(p)dp \nonumber \\
    & \approx \alpha + ({\bf w}_i^T, {\bf z}_i^T)\bm{\zeta}+ \sum_{k=1}^K \hat{Q}^*_{\mathcal{L}_ik}\gamma_k + \sum_{k=1}^K \hat{Q}^*_{\mathcal{L}_i^c,k}\eta_k , \label{eq:like}
\end{align}
\noindent where $\int_{0}^1 \hat{Q}_{\mathcal{X}_i}(p) \beta_{\mathcal{X}}(p)dp = \int \hat{Q}_{\mathcal{L}_i}(p) \beta_{\mathcal{L}}(p)dp + \int \hat{Q}_{\mathcal{L}^c_i}(p) \beta_{\mathcal{L}^c}(p)dp \approx \sum_{k=1}^K \hat{Q}^*_{\mathcal{L}_ik}\gamma_k + \sum_{k=1}^K \hat{Q}^*_{\mathcal{L}_i^c,k}\eta_k$,  $\beta(p) $ denotes the functional regression coefficient that can be represented as the basis expansion $\beta(p)\approx \sum_{k=1}^K \eta_k \psi_k(p) $, $Q^*$ are defined in (\ref{eq:basis_expansion}), ${\bf w}$ denotes the demographic features with corresponding effects denoted by ${\bm \zeta}$, and ${\bf z}$ denotes the structural radiomics features derived from the arterial phase of the T1w-MRI scans with corresponding effects $\bm{\gamma}$.  Model (\ref{eq:like}) accounts for heterogeneity by assigning separate effects for the lesional and peri-lesional areas, and via smoothed quantile curves capturing rich distributional information across all pixels. We the following penalized likelihood criteria to estimate the model parameters in (\ref{eq:like}):
\begin{eqnarray}
    \hat{\bm{\theta}} = \mbox{arg min}_{\bm{\theta}} - \log(L({\bf y}\mid \mathcal{X},\bm{\theta}) + \lambda  |\bm{\theta}|_1, \label{eq:penlike}
\end{eqnarray}
where $\bm{\theta}=(\alpha, \bm{\zeta,\gamma,\eta})$ denotes the collection of model parameters, $\lambda$ denotes the penalty parameter for the $L_1$ penalty, with a larger value indicating greater sparsity and vice-versa. The imposed $L_1$ penalty down-weights the unimportant coefficients to zero, thus reducing overfitting and resulting in model parsimony. The penalized logistic regression model in  (\ref{eq:like})-(\ref{eq:penlike}) was implemented in R using the {\it glmnet} package.  The regularization parameter $\lambda$ was selected  by using 10-fold cross-validation on the entire dataset, and this choice of $\lambda$ is subsequently used in all model fitting to report the leave-out-one cross-validation prediction errors. {\color{black}We denote the proposed model in (\ref{eq:like})-(\ref{eq:penlike}) as scalar-on-functional quantile classification model} that is used to  address the following analysis goals.

\subsubsection{Description of Analysis Goals}
We focus on three distinct aims with strong translational implications. The first goal (Aim 1: Diagnostic Aim) involves differentiating between cases and controls using the Liver EPM images at both pre-diagnostic and diagnostic visits using a cross-sectional analysis. We note that all the control subjects had one visit, whereas individuals with lesions had two visits. This resulted in a classification analysis with 97 control subjects and {\color{black}91 }images with lesions. Aim 1A uses the whole liver EPM images without knowledge of the lesion masks when computing the functional quantlet features to predict cases versus controls. 
This set-up is challenging since the information from the lesional tissue is somewhat masked by the information contained in the healthy tissues that encompasses most of the liver, especially given the relatively small lesion sizes. The resulting analysis contains limited information about the lesion characteristics that potentially yields reduced power to differentiate cases versus controls. {\color{black}We further extend the analysis to include information about the lesion masks for cases in Aim 1B. Aim 1B is expected to produce superior classification accuracy since the quantile distributions from the cases arise from lesional and peri-lesional areas only, which provides a greater differentiating power between cases and controls. } The second goal (Aim 2: Risk Stratification) focuses on lesion level classification. In particular, we are interested in differentiating between aggressive tumors vs mild tumors in a cross-sectional manner by combining information across both visits but without explicitly accounting for longitudinal dependence within subjects. A tumor is labeled as aggressive if it has grade  higher than 3, and mild if the lesion grade is 3 or lower. This results in a sample size of 84 for this analysis, with 39 tumors having grades higher than 3, and the rest having lower grades.

Transitioning beyond Aims 1-2 that are designed primarily for demonstrating the practicality of the proposed functional quantlet features, we now delve into our ultimate and pivotal goal of early detection. Aim 3 specifically concentrates on forecasting tumor progression during the diagnostic visit using imaging data solely from the pre-diagnostic visit. The primary goal is to discern which lesions are likely to remain stable over time compared to those anticipated to deteriorate, indicated by an escalation in tumor grade between visits. This analysis is conducted at the lesion level, involving 35 samples. The modest sample size presents a challenge for most models to effectively predict lesion progression, particularly when considering the large size of the images.


In order to include information from the adjoining areas surrounding the lesions that can potentially be impacted by the lesion in the future, we construct a sphere-shaped region of interest (ROI) around the lesion that encompasses the entire lesion and is approximately 2 times the volume of the corresponding lesion. These areas, referred to as peri-lesional areas, are likely to be significantly influenced by tumor progression but remain unaffected for stable tumors that exhibit minimal changes over time. Consequently, they are anticipated to contain valuable information crucial to our objective of early detection and are thus integral to our analysis. Notably, the inclusion of these adjacent peri-lesional areas resulted in considerable improvements in classification accuracy for Aim 3 compared to a more restrictive analysis focused strictly on the lesion mask (results not presented here).

\subsubsection{Benchmark approaches for comparison}
A leave-out-one cross-validation (LOOCV) strategy was used to evaluate the performance of the proposed approach versus various competing models that differed in terms of the sets of predictors used for classification under a penalized logistic regression approach. In particular, we compared with classification analyses involving: (i) demographic covariates only; (ii) summary measures such as mean or median of the pixel-level EPM values along with demographics; (iii) empirical quantiles along with demographics; (iv) summary structural radiomics features with demographics; and (v) the features considered in (ii)-(iii) along with structural radiomics features. We consider various numbers of basis coefficients corresponding to classification analysis based on quantiles. 

Several metrics were used to evaluate performance. These measures included: (a) sensitivity that is defined as the power to detect true cases in Aim 1, the power to detect tumors with higher grades in Aim 2, and the power to detect aggressive tumors that longitudinally progresses to higher grades in Aim 3; (b) the corresponding specificity rates that is equivalent to 1- false discovery rate; (c)  the F1 score that is the harmonic mean of sensitivity and specificity; and (d) accuracy that is defined as the proportion of correctly identified cases and controls. In our overall analysis, we fit the models with the extracted radiomics features as well as using principal components derived from these features that are designed to remove collinearity. However, the latter PCA based analysis were not included since they did not produce improvements.



\section{Results}


The results for Aims 1-3 are presented in Tables \ref{tab:aim1}-\ref{tab:aim3}, respectively. {\color{black}We report the results over varying number of basis coefficients included in the classification model, where these coefficients are included based on their ranking (in a decreasing order of importance) in terms of their LOOCCC score. }
From Table \ref{tab:aim1}, it is clear that the best classification accuracy is obtained when using the functional quantlet features along with the structural radiomics features and demographics. For the case vs classification model, 
we observe that combining the functional quantlet features with the structural radiomics features results in considerable gains in sensitivity along with comparable or improved specificity compared to using either type of features alone. Moreover, classification based on quantlet features along with structural radiomics features results in improvements in sensitivity compared to classification based on empirical quantiles as well as summary mean(median) EPM values, when they are used in isolation or coupled with demographics.

\begin{table}[!h]
    \centering
    \begin{tabular}{|c|c|c|c|c|c|c|c|c|}
    \toprule
        & \multicolumn{8}{|c|}{Cases vs Control Diagnosis} \\
        & \multicolumn{4}{|c|}{Aim 1A} & \multicolumn{4}{|c|}{Aim 1B}\\
    \midrule
         Features & Sens & Spec & F1 & Acc & Sens & Spec & F1 & Acc \\ 
        \midrule
        D & 0.07 & 0.94 & 0.12 & 0.57 & 0.07 & 0.94 & 0.12 & 0.57 \\ 
        Mean & 0.36 & 0.76 & 0.43 & 0.59 & 0.51 & 0.90 & 0.62 & 0.73 \\ 
        Median & 0.29 & 0.80 & 0.38 & 0.58 & 0.63 & 0.93 & 0.73 & 0.80 \\ 
        \midrule
        EQ(10) & 0.26 & 0.95 & 0.40 & 0.65 & 0.63 & 0.98 & 0.76 & 0.83 \\ 
        EQ(20) & 0.26 & 0.93 & 0.39 & 0.64 & 0.64 & 0.99 & 0.77 & 0.84 \\ 
        EQ(30) & 0.29 & 0.93 & 0.42 & 0.65 & 0.68 & 1.00 & 0.81 & 0.86 \\ 
        EQ(50) & 0.29 & 0.92 & 0.42 & 0.65 & 0.70 & 1.00 & 0.83 & 0.88 \\
        \midrule
        Q(10) & 0.38 & 0.79 & 0.45 & 0.61 & 0.89 & 0.93 & 0.90 & 0.91 \\ 
        Q(20) & 0.42 & 0.76 & 0.48 & 0.61 & 0.89 & 0.92 & 0.89 & 0.90 \\ 
        Q(30) & 0.44 & 0.74 & 0.50 & 0.61 & 0.89 & 0.93 & 0.90 & 0.91 \\ 
        \midrule
        R & 0.75 & 0.94 & 0.82 & 0.86 & 0.75 & 0.94 & 0.82 & 0.86 \\ 
        \midrule
        Mean+R & 0.75 & 0.94 & 0.82 & 0.86 & 0.82 & 0.96 & 0.87 & 0.90 \\ 
        Median+R & 0.75 & 0.94 & 0.82 & 0.86 & 0.82 & 0.96 & 0.87 & 0.90 \\
        \midrule
        EQ(10)+R & 0.71 & 0.97 & 0.81 & 0.86 & 0.76 & 0.98 & 0.85 & 0.89 \\ 
        EQ(20)+R & 0.71 & 0.97 & 0.81 & 0.86 & 0.76 & 0.98 & 0.85 & 0.89 \\ 
        EQ(30)+R & 0.71 & 0.97 & 0.81 & 0.86 & 0.75 & 0.99 & 0.85 & 0.89\\ 
        EQ(50)+R & 0.71 & 0.97 & 0.81 & 0.86 & 0.79 & 0.99 & 0.88 & 0.90\\
        \midrule
        Q(10)+R & 0.78 & 0.95 & 0.84 & 0.88 & 0.88 & 0.98 & 0.92 & 0.93\\ 
        Q(20)+R & 0.74 & 0.97 & 0.83 & 0.87 & 0.88 & 0.98 & 0.92 & 0.93 \\ 
        Q(30)+R & 0.83 & 0.93 & 0.86 & 0.89 & 0.86 & 0.98 & 0.91 & 0.93\\ 
        \bottomrule 
    \end{tabular}
    \caption{Aim 1 Results: Leave-out-one CV results for case/control classification. In Aim 1A, EPM features are taken from liver masks for control and cases, whereas for Aim 1B, EPM features for the cases are taken from their lesion masks. Different predictors were considered including Mean and Median EPM, EPM empirical quantiles (EQ), EPM quantlets (Q), and PyRadiomics features (R). We considered varying number of quantiles and 70 PyRadiomics features (R).}
    \label{tab:aim1}
\end{table}

\begin{table}[!h]
    \centering
    \begin{tabular}{|c|c|c|c|c|}
    \toprule
        & \multicolumn{4}{|c|}{Risk Stratification} \\
    \midrule
         Features & Sens & Spec & F1 & Acc \\ 
        \midrule
        D & 0.55 & 0.58 & 0.58 & 0.56  \\ 
        Mean & 0.63 & 0.27 & 0.56 & 0.47  \\ 
        Median & 0.65 & 0.30 & 0.58 & 0.49  \\ 
        \midrule
        EQ(10) & 0.63 & 0.36 & 0.58 & 0.51  \\ 
        EQ(30) & 0.65 & 0.27 & 0.58 & 0.48  \\ 
        EQ(50) & 0.60 & 0.30 & 0.55 & 0.47  \\
        \midrule
        Q(10) & 0.68 & 0.36 & 0.61 & 0.53  \\ 
        Q(30) & 0.60 & 0.52 & 0.60 & 0.56  \\ 
        Q(50) & 0.83 & 0.58 & 0.76 & 0.71  \\ 
        \midrule
        R & 0.80 & 0.64 & 0.76 & 0.73  \\ 
        \midrule
        Mean+R & 0.80 & 0.64 & 0.76 & 0.73  \\ 
        Median+R &  0.80 & 0.64 & 0.76 & 0.73  \\
        \midrule
        EQ(10)+R & 0.78 & 0.67 & 0.76 & 0.73  \\ 
        EQ(30)+R & 0.78 & 0.67 & 0.76 & 0.73  \\ 
        EQ(50)+R & 0.78 & 0.67 & 0.76 & 0.73  \\
        \midrule
        Q(10)+R & 0.83 & 0.64 & 0.77 & 0.74  \\ 
        Q(30)+R & 0.85 & 0.64 & 0.79 & 0.75  \\ 
        Q(50)+R & 0.80 & 0.76 & 0.80 & 0.78  \\ 
        \bottomrule
    \end{tabular}
    \caption{Aim 2 Results: Leave-out-one CV results for cross-sectional classification of lesion grade.}  
    \label{tab:aim2}
\end{table}

{\color{black}Table \ref{tab:aim1} shows that while the sensitivity is greater than 80\%  and specificity is higher than 90\% under both Aims 1A-1B, the performance is superior for Aim 1B. The improved performance for Aim 1B is due to the fact that the quantile distributions arising from the lesional and peri-lesional areas were considerably more different than the quantile distributions arising from healthy liver parenchyma. In contrast, the entire liver image was used for Aim 1A without any lesion level information that may have diluted the discriminative ability of the methods.} 
For Aim 2 involving lesion grade classification (see Table \ref{tab:aim2}), similar trends hold. In particular, combining the functional quantlet features with the structural radiomics features results in sensitivity of 80\% that is higher than using each type of features alone in the model, and specificity of 76\% with 50 quantlets. In contrast, using the   empirical quantiles, as well as summary measures (mean,median), results in lower sensitivity and specificity, when used alone or in conjunction with structural radiomics features. {\color{black}This points to the limited utility of these features in risk stratification.} 

\begin{table}[!h]
    \centering
    \begin{tabular}{|c|c|c|c|c|c|c|c|c|}
    \toprule
        & \multicolumn{4}{|c|}{Progression (lesion)} & \multicolumn{4}{|c|}{Progression (ROI)} \\
    \midrule
         Features & Sens & Spec & F1 & Acc & Sens & Spec & F1 & Acc \\ 
        \midrule
        D & 0.18 & 0.90 & 0.27 & 0.65 & 0.18 & 0.90 & 0.27 & 0.65 \\ 
        Mean & 0 & 0.83 & 0 & 0.54 & 0.18 & 0.85 & 0.25 & 0.61 \\ 
        Median & 0.10 & 0.89 & 0.15 & 0.61  & 0.27 & 0.80 & 0.33 & 0.61 \\ 
        \midrule
        EQ(10) & 0 & 0.94 & 0 & 0.61 & 0.46 & 0.70 & 0.46 & 0.61 \\ 
         EQ(20) & 0.10 & 0.94 & 0.17 & 0.64 & 0.46 & 0.80 & 0.50 & 0.68 \\ 
          EQ(50) & 0 & 0.89 & 0 & 0.57  & 0.46 & 0.80 & 0.50 & 0.68 \\
        \midrule
        Q(10) & 0 & 0.89 & 0 & 0.57  & 0.46 & 1.00 & 0.63 & 0.81 \\ 
        Q(20) & 0.20 & 0.83 & 0.27 & 0.61 & 0.46 & 1.00 & 0.63 & 0.81 \\ 
         Q(50) & 0.50 & 0.94 & 0.62 & 0.79 & 0.27 & 0.95 & 0.40 & 0.71 \\ 
        \midrule
        R & 0.60 & 0.94 & 0.71 & 0.82  & 0.64 & 0.90 & 0.70 & 0.81 \\ 
        \midrule
       Mean+R & 0.60 & 0.94 & 0.71 & 0.82  & 0.64 & 0.80 & 0.64 & 0.74 \\ 
        Median+R & 0.60 & 0.94 & 0.71 & 0.82  & 0.64 & 0.90 & 0.70 & 0.81 \\
        \midrule
        EQ(10)+R & 0.60 & 1 & 0.75 & 0.86 & 0.64 & 0.90 & 0.70 & 0.81 \\ 
        EQ(20)+R & 0.60 & 1 & 0.75 & 0.86 & 0.64 & 0.90 & 0.70 & 0.81 \\ 
        \midrule
       Q(10)+R & 0.80 & 0.94 & 0.84 & 0.90  & 0.82 & 0.85 & 0.78 & 0.84 \\ 
        Q(20)+R & 0.80 & 0.94 & 0.84 & 0.90  & 0.82 & 0.85 & 0.78 & 0.84 \\ 
        \bottomrule
    \end{tabular}
    \caption{Aim 3 Results: Leave-out-one CV results for change in lesion grade across visits ($>1$ vs. $\leq 1$). Two separate analysis are performed, one exclusively involving the lesion voxels, and the other involving the ROI that contains additional peri-lesional tissues surrounding the lesion. } 
    \label{tab:aim3}
\end{table}

For predicting tumor progression (Aim 3), we 
report results from two separate analysis in Table \ref{tab:aim3}. In the first analysis the prediction is based on pixels lying within the lesion exclusively (denoted as lesional analysis), whereas for the second analysis, the prediction is based on pre-defined ROIs that contain some peri-lesional areas surrounding the lesion (denoted as ROI analysis). We find that the best results are obtained under the proposed approach combining the functional quantlet and structural radiomics features derived from the lesions directly, without including peri-lesional areas. In particular, the analysis focusing on the lesional areas exclusively produces comparable sensitivity (80\%) but much higher specificity (94\%) under the proposed approach when compared to the corresponding ROI level analysis involving additional peri-lesional areas. We therefore conclude that including peri-lesional areas surrounding the lesion leads to loss of specificity (or equivalently an increase in the false discovery rate) without providing any obvious benefits in terms of an increase in sensitivity. Moreover, alternate analysis involving empirical quantiles and summary measures produce considerably lesser sensitivity and comparable specificity under both types of analysis (lesional analysis as well as ROI analysis).  Finally, the F-1 score is markedly higher for the lesional analysis under the proposed approach. Coupled with high sensitivity levels, these results highlight the prognostic value of the imaging biomarkers. 


Overall, for all the three aims, a combination of the functional quantlet features and structural radiomics features produced the best results with increased sensitivity and comparable specificity that translated to improved accuracy and F-1 scores.  The improvements under the proposed smoothed quantile approach over empirical quantiles is more noticeable for Aims 2 and 3 where the lesion mask information was incorporated. Moreover, combining the functional quantlet features with structural radiomics produces greater improvements in classification compared to the combination of empirical quantiles and structural radiomics features. This suggests that the smoothed quantile distribution can leverage informative patterns from the lesion for classification when the lesion location is specified, and that they are more amenable for integrative analysis in combination with  structural imaging features.  
For Aim 3, the sharp increase in sensitivity for the lesional analysis points to the ability of the proposed approach (combining functional quantlet and structural radiomic features) to detect lesions that are likely to grow aggressively over time. In contrast, the poor sensitivity of the routinely used summary measures (mean, median) is not desirable and points to the limits of existing analysis approaches in terms of accommodating heterogeneity across the image. 


\section{Discussion}
In this paper, we proposed a novel DDA based approach for image-based prediction that accounts for image heterogeneity and incorporates information from the pixel-level quantile distribution as functional covariates, along with summary structural radiomics features. Our extensive numerical analysis revealed that the functional quantlet approach yielded improvements in terms of out-of-sample classification, owing to the ability to capture minute distributional differences often ignored by classical approaches. 
The overarching framework developed here for image based risk prediction and early diagnosis is expected to have considerable impact in cancer imaging. Some additional considerations are elaborated below.


\vskip 10pt

{\noindent \underline{Statistical Consideration and Potential Pitfalls:}}
For the lesion-level classification, we attempted to utilize radiomic information only within or directly surrounding each lesion for cross-sectional and longitudinal prediction. However, one challenge that arises is the vast amount of heterogeneity in the size (i.e., number of pixels) comprising the different lesions. In this sense, our analysis was limited with respect to smaller lesions which did not contain as much information. We note potential bias arising from the fact that lower severity lesions tended to have smaller sizes. Furthermore, these smaller lesions limited the scope of our feature space both for the empirical quantiles, and the quantlet coefficients, which were estimated over a unified grid, to limit cost and avoid additional issues.

In addition to the lesion size differences, we note that the choice of ROIs representing the peri-lesional areas may impact the final results. In our analysis, we selected ROIs as
 spherical regions centered around the lesion centroid, with approximately 2 times the volume of the lesion. However, the sensitivity of the approach to the choice of size and shape of these ROIs need to be investigated further. 
 Moreover, the analysis in Aim 3 demonstrated that the inclusion of surrounding peri-lesional tissues is not always guaranteed to produce improvements compared to an analysis focused exclusively on lesional pixels. Ultimately, the performance may depend on additional considerations such as sample size.
{\color{black}Finally, the recommended number of quantlets to be included in the model should be such that the proposed method achieves at least 80\% sensitivity and specificity in a validation sample.}


\vskip 10pt

{\noindent \underline{Computational Scalability:}} The computational scalability of functional quantlet classification relies largely on the process of estimating quantlet coefficients. 
The computational speed of the quantlet extraction process depends on factors such as the variability in image sizes. When the image sizes vary widely, the quality of the empirical quantiles computed from the smaller images may be compromised resulting in computation challenges related to issues with 
non-uniform sampling, Gram-Schmidt orthogonalization, and potential memory issues in the basis dictionary construction. To get around these challenges, we computed quantlets based on $512$ empirical quantiles for the case-control classification (entire liver) and $128$ empirical quantiles for the lesion-level classifications. While this downsampling may result is minimal information loss, it facilitates practical implementation and stability.

\vskip 10pt

{\noindent \underline{Future Directions:}} The framework developed in this article can be generalized in several different directions. For example, one could evaluate whether longitudinal changes in the EPM images are associated with tumor progression. The proposed framework can also be used to predict response to therapy in future visits, in clinical trials. Future analysis could also be enriched by including additional non-imaging biomarkers such as blood based biomarkers frequently used in HCC. Finally, the proposed DDA methodology can be broadly applied to other types of imaging modalities and disease areas. 

\printbibliography[heading=bibintoc,title={References}]

@article{tzartzeva2018surveillance,
  title={Surveillance imaging and alpha fetoprotein for early detection of hepatocellular carcinoma in patients with cirrhosis: a meta-analysis},
  author={Tzartzeva, Kristina and Obi, Joseph and Rich, Nicole E and Parikh, Neehar D and Marrero, Jorge A and Yopp, Adam and Waljee, Akbar K and Singal, Amit G},
  journal={Gastroenterology},
  volume={154},
  number={6},
  pages={1706--1718},
  year={2018},
  publisher={Elsevier}
}

@article{smith2007spatial,
  title={Spatial Bayesian variable selection with application to functional magnetic resonance imaging},
  author={Smith, Michael and Fahrmeir, Ludwig},
  journal={J. Am. Stat. Assoc.},
  volume={102},
  number={478},
  pages={417--431},
  year={2007},
  publisher={Taylor \& Francis}
}

@article{lambin2012radiomics,
  title={Radiomics: extracting more information from medical images using advanced feature analysis},
  author={Lambin, Philippe and Rios-Velazquez, Emmanuel and Leijenaar, Ralph and Carvalho, Sara and Van Stiphout, Ruud GPM and Granton, Patrick and Zegers, Catharina ML and Gillies, Robert and Boellard, Ronald and Dekker, Andr{\'e} and others},
  journal={Euro. J. of Cancer},
  volume={48},
  number={4},
  pages={441--446},
  year={2012},
  publisher={Elsevier}
}

@article{lambin2017radiomics,
  title={Radiomics: the bridge between medical imaging and personalized medicine},
  author={Lambin, Philippe and Leijenaar, Ralph TH and Deist, Timo M and Peerlings, Jurgen and De Jong, Evelyn EC and Van Timmeren, Janita and Sanduleanu, Sebastian and Larue, Ruben THM and Even, Aniek JG and Jochems, Arthur and others},
  journal={Nat Rev Clin Onc},
  volume={14},
  number={12},
  pages={749--762},
  year={2017},
  publisher={Nature Publishing Group UK London}
}

@article{aerts2014decoding,
  title={Decoding tumour phenotype by noninvasive imaging using a quantitative radiomics approach},
  author={Aerts, Hugo JWL and Velazquez, Emmanuel Rios and Leijenaar, Ralph TH and Parmar, Chintan and Grossmann, Patrick and Carvalho, Sara and Bussink, Johan and Monshouwer, Ren{\'e} and Haibe-Kains, Benjamin and Rietveld, Derek and others},
  journal={Nat Comm},
  volume={5},
  pages={4006},
  year={2014},
  publisher={Nature Publishing Group UK London}
}

@article{larue2017quantitative,
  title={Quantitative radiomics studies for tissue characterization: a review of technology and methodological procedures},
  author={Larue, Ruben THM and Defraene, Gilles and De Ruysscher, Dirk and Lambin, Philippe and Van Elmpt, Wouter},
  journal={The British J of Rad},
  volume={90},
  number={1070},
  pages={20160665},
  year={2017},
  publisher={The British Institute of Radiology.}
}

@article{dagogo2018tumour,
  title={Tumour heterogeneity and resistance to cancer therapies},
  author={Dagogo-Jack, Ibiayi and Shaw, Alice T},
  journal={Nat Rev Clin Onc},
  volume={15},
  number={2},
  pages={81--94},
  year={2018},
  publisher={Nature Publishing Group UK London}
}

@article{ramsay1991some,
  title={Some tools for functional data analysis},
  author={Ramsay, James O and Dalzell, CJ1125714},
  journal={J. of the Royal Stat Soc Series B},
  volume={53},
  number={3},
  pages={539--561},
  year={1991},
  publisher={Oxford University Press}
}

@article{morris2015functional,
  title={Functional regression},
  author={Morris, Jeffrey S},
  journal={Ann. Rev Stat and Its Appl},
  volume={2},
  pages={321--359},
  year={2015},
  publisher={Annual Reviews}
}

@article{ma2022multi,
  title={Multi-task learning with high-dimensional noisy images},
  author={Ma, Xin and Kundu, Suprateek and Alzheimer’s Disease Neuroimaging Initiative},
  journal={J Am Stat Assoc},
  pages={1--14},
  year={2022},
  publisher={Taylor \& Francis}
}

@article{zipunnikov2011functional,
  title={Functional principal component model for high-dimensional brain imaging},
  author={Zipunnikov, Vadim and Caffo, Brian and Yousem, David M and Davatzikos, Christos and Schwartz, Brian S and Crainiceanu, Ciprian},
  journal={NeuroImage},
  volume={58},
  pages={772--784},
  year={2011},
  publisher={Elsevier}
}

@article{wang2017generalized,
  title={Generalized scalar-on-image regression models via total variation},
  author={Wang, Xiao and Zhu, Hongtu and Alzheimer’s Disease Neuroimaging Initiative},
  journal={J Am Stat Assoc},
  volume={112},
  number={519},
  pages={1156--1168},
  year={2017},
  publisher={Taylor \& Francis}
}

@article{reiss2015wavelet,
  title={Wavelet-domain regression and predictive inference in psychiatric neuroimaging},
  author={Reiss, Philip T and Huo, Lan and Zhao, Yihong and Kelly, Clare and Ogden, R Todd},
  journal={The Ann of Appl Stat},
  volume={9},
  number={2},
  pages={1076},
  year={2015},
  publisher={NIH Public Access}
}

@article{augustin2017modelling,
  title={Modelling a response as a function of high-frequency count data: the association between physical activity and fat mass},
  author={Augustin, Nicole H and Mattocks, Calum and Faraway, Julian J and Greven, Sonja and Ness, Andy R},
  journal={Stat Meth in Med Res},
  volume={26},
  number={5},
  pages={2210--2226},
  year={2017},
  publisher={Sage Publications Sage UK: London, England}
}

@article{dumuid2020compositional,
  title={Compositional data analysis in time-use epidemiology: what, why, how},
  author={Dumuid, Dorothea and Pedi{\v{s}}i{\'c}, {\v{Z}}eljko and Palarea-Albaladejo, Javier and Mart{\'\i}n-Fern{\'a}ndez, Josep Antoni and Hron, Karel and Olds, Timothy},
  journal={Int J of Env Res and Pub Health},
  volume={17},
  number={7},
  pages={2220},
  year={2020},
  publisher={MDPI}
}

@article{ghodrati2022distribution,
  title={Distribution-on-distribution regression via optimal transport maps},
  author={Ghodrati, Laya and Panaretos, Victor M},
  journal={Biometrika},
  volume={109},
  number={4},
  pages={957--974},
  year={2022},
  publisher={Oxford University Press}
}

@article{talska2021compositional,
  title={Compositional scalar-on-function regression with application to sediment particle size distributions},
  author={Talsk{\'a}, Ren{\'a}ta and Hron, Karel and Grygar, Tom{\'a}{\v{s}} Matys},
  journal={Math Geosciences},
  volume={53},
  number={7},
  pages={1667--1695},
  year={2021},
  publisher={Springer}
}

@article{matabuena2021distributional,
  title={Distributional data analysis with accelerometer data in a nhanes database with nonparametric survey regression models.},
  author={Matabuena, M and Petersen, A},
  journal={arXiv:2104.01165},
  year={2021}
}

@software{MATLAB,
year = {2022},
author = {The MathWorks Inc.},
title = {MATLAB version: 9.13.0 (R2022b)},
publisher = {The MathWorks Inc.},
address = {Natick, Massachusetts, United States},
url = {https://www.mathworks.com}
}

@article{mahmud2020risk,
  title={Risk Factors and Center-Level Variation in Hepatocellular Carcinoma Under-Staging for Liver Transplantation},
  author={Mahmud, Nadim and Hoteit, Maarouf A and Goldberg, David S},
  journal={Liver Trans},
  volume={26},
  number={8},
  pages={977--988},
  year={2020},
  publisher={Wiley Online Library}
}

@article{chernyak2018liver,
  title={Liver Imaging Reporting and Data System (LI-RADS) version 2018: imaging of hepatocellular carcinoma in at-risk patients},
  author={Chernyak, Victoria and Fowler, Kathryn J and Kamaya, Aya and Kielar, Ania Z and Elsayes, Khaled M and Bashir, Mustafa R and Kono, Yuko and Do, Richard K and Mitchell, Donald G and Singal, Amit G and others},
  journal={Radiology},
  volume={289},
  number={3},
  pages={816--830},
  year={2018},
  publisher={Radiological Society of North America}
}

@article{pan2021safe,
  title={A Safe Feature Elimination Rule for $ L\_ $\{$1$\}$ $ L 1-Regularized Logistic Regression},
  author={Pan, Xianli and Xu, Yitian},
  journal={IEEE Trans on Pat Anal and Mach Intel},
  volume={44},
  number={9},
  pages={4544--4554},
  year={2021},
  publisher={IEEE}
}

@article{donoho1995wavelet,
  title={Wavelet shrinkage: asymptopia?},
  author={Donoho, David L and Johnstone, Iain M and Kerkyacharian, G{\'e}rard and Picard, Dominique},
  journal={J of the Royal Stat Soc: Series B },
  volume={57},
  number={2},
  pages={301--337},
  year={1995},
  publisher={Wiley Online Library}
}

@inproceedings{nikzad2022characterization,
  title={Characterization of the Imaging Signature of Hepatocellular Carcinoma with Enhancement Pattern Mapping},
  author={Nikzad, Newsha and Fuentes, D and Roach, Millicent and Chowdhury, Tasadduk and Cagley, Matthew and Badawy, Mohamed and Hassan, Manal M and Elsayes, K and Beretta, Laura and Koay, Eugene J and others},
  booktitle={HEPATOLOGY},
  volume={76},
  pages={S1356--S1356},
  year={2022},
  organization={WILEY 111 RIVER ST, NJ USA}
}

@article{van2017computational,
  title={Computational radiomics system to decode the radiographic phenotype},
  author={Van Griethuysen, Joost JM and Fedorov, Andriy and Parmar, Chintan and Hosny, Ahmed and Aucoin, Nicole and Narayan, Vivek and Beets-Tan, Regina GH and Fillion-Robin, Jean-Christophe and Pieper, Steve and Aerts, Hugo JWL},
  journal={Can Res},
  volume={77},
  number={21},
  pages={e104--e107},
  year={2017},
  publisher={AACR}
}

@article{feng2020bayesian,
  title={Bayesian scalar on image regression with nonignorable nonresponse},
  author={Feng, Xiangnan and Li, Tengfei and Song, Xinyuan and Zhu, Hongtu},
  journal={J Am Stat Assoc},
  volume={115},
  number={532},
  pages={1574--1597},
  year={2020},
  publisher={Taylor \& Francis}
}

@article{guhaniyogi2017bayesian,
  title={Bayesian tensor regression},
  author={Guhaniyogi, Rajarshi and Qamar, Shaan and Dunson, David B},
  journal={J Mach Learn Res},
  volume={18},
  number={1},
  pages={2733--2763},
  year={2017},
  publisher={JMLR. org}
}

@article{kundu2023bayesian,
  title={Bayesian longitudinal tensor response regression for modeling neuroplasticity},
  author={Kundu, Suprateek and Reinhardt, Alec and Song, Serena and Han, Joo and Meadows, M Lawson and Crosson, Bruce and Krishnamurthy, Venkatagiri},
  journal={Human Brain Mapping},
  year={2023},
  publisher={Wiley Online Library}
}

@article{lee2014spatial,
  title={Spatial Bayesian variable selection models on functional magnetic resonance imaging time-series data},
  author={Lee, Kuo-Jung and Jones, Galin L and Caffo, Brian S and Bassett, Susan Spear},
  journal={Bayesian Analysis},
  volume={9},
  number={3},
  pages={699},
  year={2014},
  publisher={NIH Public Access}
}

@article{friston1995characterizing,
  title={Characterizing evoked hemodynamics with fMRI},
  author={Friston, Karl J and Frith, Chris D and Turner, Robert and Frackowiak, Richard SJ},
  journal={Neuroimage},
  volume={2},
  number={2},
  pages={157--165},
  year={1995},
  publisher={Elsevier}
}

@article{hosking1990moments,
  title={L-moments: analysis and estimation of distributions using linear combinations of order statistics},
  author={Hosking, Jonathan RM},
  journal={J of the Royal Stat Soc, Series B},
  volume={52},
  number={1},
  pages={105--124},
  year={1990},
  publisher={Oxford University Press}
}

@article{ghosal2023distributional,
  title={Distributional data analysis via quantile functions and its application to modeling digital biomarkers of gait in Alzheimer’s disease},
  author={Ghosal, Rahul and Varma, Vijay R and Volfson, Dmitri and Hillel, Inbar and Urbanek, Jacek and Hausdorff, Jeffrey M and Watts, Amber and Zipunnikov, Vadim},
  journal={Biostatistics},
  volume={24},
  number={3},
  pages={539--561},
  year={2023},
  publisher={Oxford University Press}
}

@article{park2020enhancement,
  title={Enhancement pattern mapping technique for improving contrast-to-noise ratios and detectability of hepatobiliary tumors on multiphase computed tomography},
  author={Park, Peter C and Choi, Gye W and M. Zaid, Mohamed and Elganainy, Dalia and Smani, Danyal A and Tomich, John and Samaniego, Ray and Ma, Jingfei and Tamm, Eric P and Beddar, Sam and others},
  journal={Med Phys},
  volume={47},
  number={1},
  pages={64--74},
  year={2020},
  publisher={Wiley Online Library}
}

@article{yildirim2023advances,
  title={Advances in the Early Detection of Hepatobiliary Cancers},
  author={Y{\i}ld{\i}r{\i}m, Hasan {\c{C}}a{\u{g}}r{\i} and Kavgaci, Gozde and Chalabiyev, Elvin and Dizdar, Omer},
  journal={Cancers},
  volume={15},
  number={15},
  pages={3880},
  year={2023},
  publisher={MDPI}
}

@article{tibshirani1996regression,
  title={Regression shrinkage and selection via the lasso},
  author={Tibshirani, Robert},
  journal={J of the Roy Stat Soc: Series B},
  volume={58},
  number={1},
  pages={267--288},
  year={1996},
  publisher={Wiley Online Library}
}

@article{yang2020quantile,
  title={Quantile function on scalar regression analysis for distributional data},
  author={Yang, Hojin and Baladandayuthapani, Veerabhadran and Rao, Arvind UK and Morris, Jeffrey S},
  journal={J Am Stat Assoc},
  volume={115},
  number={529},
  pages={90--106},
  year={2020},
  publisher={Taylor \& Francis}
}

@INPROCEEDINGS{Avulamean,
  author={Avula, Madhuri and Lakkakula, Narasimha Prasad and Raja, Murali Prasad},
  booktitle={2014 8th Asia Modelling Symposium}, 
  title={Bone Cancer Detection from MRI Scan Imagery Using Mean Pixel Intensity}, 
  year={2014},
  volume={},
  number={},
  pages={141-146},
  doi={10.1109/AMS.2014.36}}

\end{document}